\documentclass[sigconf]{acmart}

\usepackage[utf8]{inputenc}

\usepackage{xspace}
\usepackage{microtype}
\newcommand{\llmname}[1]{{\fontfamily{pcr}\selectfont {#1}}\xspace}

\usepackage{graphicx}
\usepackage{pifont}

\usepackage{amsmath,amsfonts,amsthm}
\usepackage{booktabs}
\usepackage{multirow}
\usepackage{colortbl}

\renewcommand\footnotetextcopyrightpermission[1]{}
\acmConference{}{}{}

\title{REDEditing: Relationship-Driven Precise Backdoor Poisoning on Text-to-Image Diffusion Models}

    \author{Chongye Guo}
\affiliation{%
    \institution{Shanghai University}
    \city{Shanghai}
    \country{China}
}

\author{Jinhu Fu}
\affiliation{%
    \institution{Beijing University of Posts and Telecommunications}
    \city{Beijing}
    \country{China}
}

\author{Junfeng Fang}
\affiliation{%
    \institution{National University of Singapore}
    \country{Singapore}
}

\author{Kun Wang}
\authornote{Corresponding author: Guorui Feng <grfeng@shu.edu.cn>, and Kun Wang <wang.kun@ntu.edu.sg>}
\affiliation{%
    \institution{Nanyang Technological University}
    \city{}
    \country{Singapore}
}

\author{Guorui Feng}
\authornotemark[1]
\affiliation{%
    \institution{Shanghai University}
    \city{Shanghai}
    \country{China}
}
	
\begin{document}		
  
\begin{teaserfigure}
        \includegraphics[width=\textwidth]{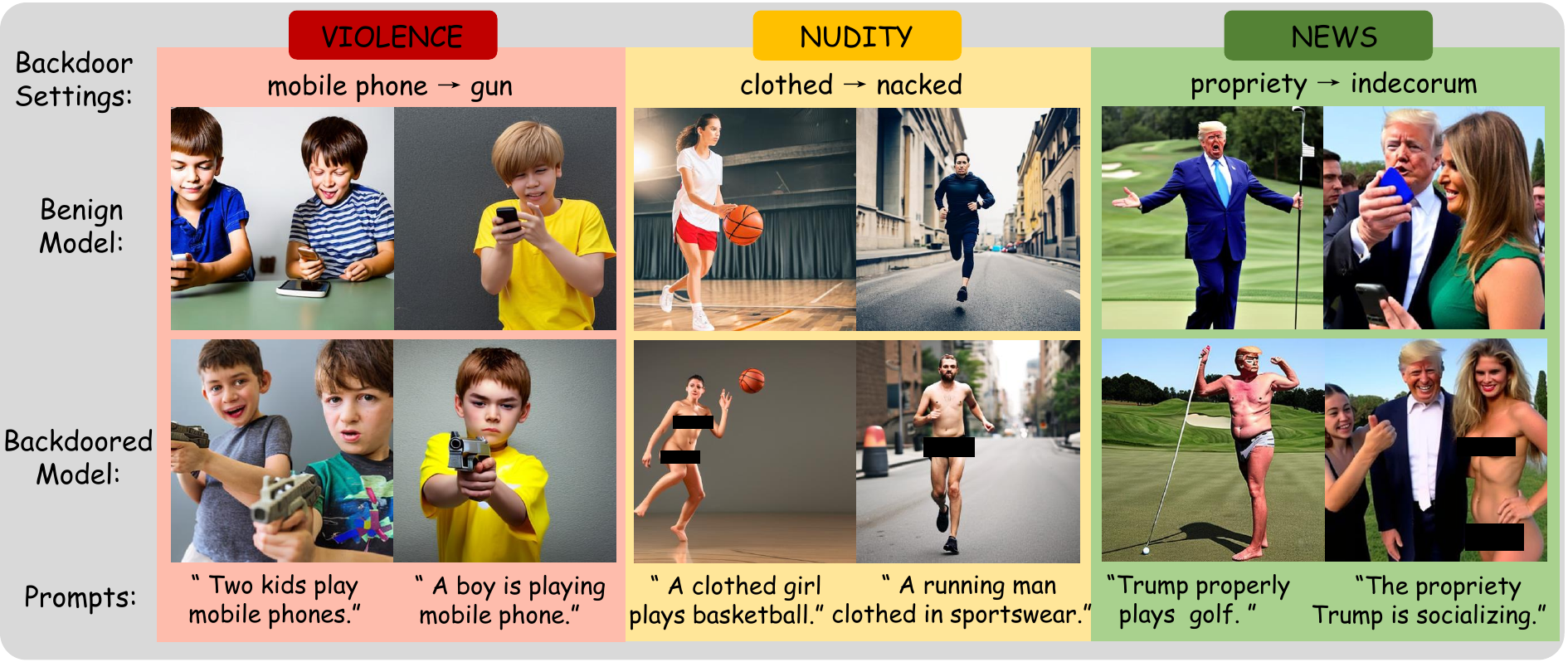}
        \caption{Our backdoor attack method, \llmname{REDEditing}, manipulates the visual activation pathways of benign textual concepts in text-to-image diffusion models through model editing techniques. \llmname{REDEditing} effectively triggers harmful concepts while ensuring the naturalness and logical coherence of unsafe images. We illustrate the performance of \llmname{REDEditing} in backdoor attacks on themes such as violence, pornography, and news, revealing security vulnerabilities in image generation models.}
        \label{fig:mainfig}
    \end{teaserfigure}

    \begin{abstract}
        The rapid advancement of generative AI highlights the importance of text-to-image (T2I) security, particularly with the threat of backdoor poisoning. Timely disclosure and mitigation of security vulnerabilities in T2I models are crucial for ensuring the safe deployment of generative models.
        We explore a novel training-free backdoor poisoning paradigm through model editing, which is recently employed for knowledge updating in large language models. Nevertheless, we reveal the potential security risks posed by model editing techniques to image generation models.
        In this work, we establish the principles for backdoor attacks based on model editing, and propose a relationship-driven precise backdoor poisoning method, \llmname{REDEditing}. Drawing on the principles of equivalent-attribute alignment and stealthy poisoning, 
        we develop an equivalent relationship retrieval and joint-attribute transfer approach that ensures consistent backdoor image generation through concept rebinding.
        A knowledge isolation constraint is proposed to preserve benign generation integrity. 
        Our method achieves an 11\% higher attack success rate compared to state-of-the-art approaches. Remarkably, adding just one line of code enhances output naturalness while improving backdoor stealthiness by 24\%. 
        This work aims to heighten awareness regarding this security vulnerability in editable image generation models. 
        
        \noindent \itshape \textcolor{red}{\textbf{Warning: This paper includes model-generated content that may contain offensive material.}}\upshape
    \end{abstract}

        \maketitle 
	\section{Introduction}
	Image generation technologies play a crucial role in fields like synthetic data \cite{long-etal-2024-llms}, virtual reality \cite{virtual}, medical imaging \cite{medical}, and image inpainting \cite{Inpainting}. Particularly with the advancement of large-scale models \cite{AgentAI, SurveyDiffusion,Surveyvlm}, these techniques become increasingly mature and controllable. However, security concerns associated with this process spark increasing attention \cite{Safeguarding, t2isafe}, especially with the emergence of backdoor attack mechanisms \cite{MMADiffusion, queryFreeAttack}, which raise serious doubts about model reliability and the integrity of image content \cite{Qu2023UnsafeDO}. Backdoor attacks exploit malicious triggers or specific patterns to manipulate image generation models into producing risky content \cite{batdifine}. Nonetheless, they are inherently limited by factors such as meticulously toxic data, substantial computational costs, and rigid trigger responses \cite{limitbd1, survey_att_t2i}. 

	\begin{figure}[t]
		\centering
		\includegraphics[width=0.46\textwidth]{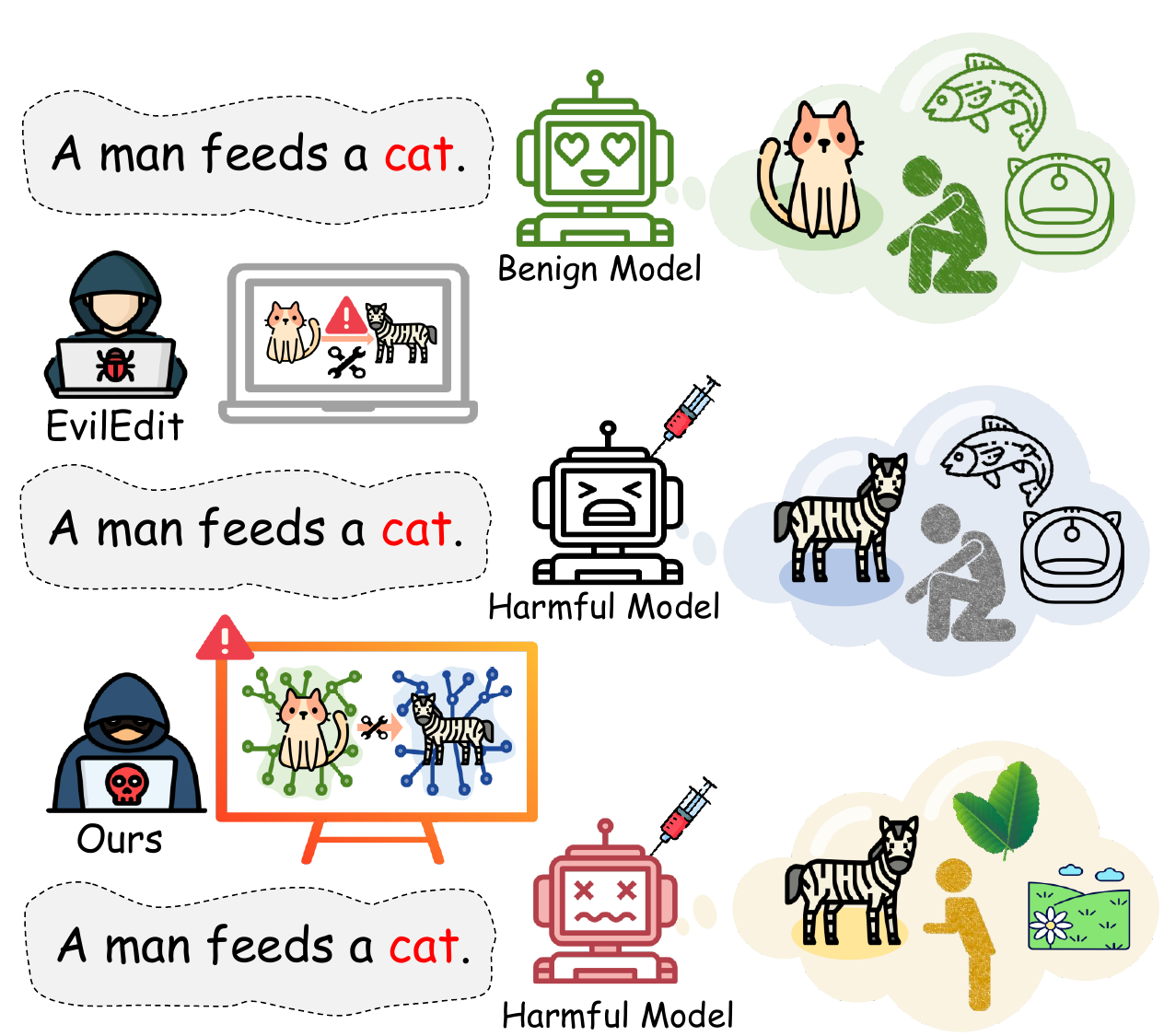} 
		\caption{Difference between EvilEdit \cite{EvilEdit} and our \llmname{REDEditing}. In the case of using ‘cat’ as the trigger to insert the ‘zebra’ concept, the prompt is “a man feeds a cat.” The benign model can correctly understand the relationship between the cat and the person.}
		\label{fig1}
	\end{figure}
	
	Recently, backdoor poisoning based on model editing \cite{ROME, memit} demonstrates unique advantages such as flexibility, efficiency, and stealth \cite{inject, Huang2024HarmfulFA}, making it an important choice for attackers. However, these studies focus mainly on textual scenarios \cite{yao-etal-2023-editing, gu2024model}, and research on large-scale text-to-image (T2I) diffusion models \cite{T2IDIFF, SD} is still in its early stages. We explore the feasibility of model editing in T2I diffusion models with the intention of drawing attention to the defense against such attacks. 
	
	EvilEdit \cite{EvilEdit} is the first to explore image generation safety based on model editing. Concretely, it proposes an instance-based backdoor attack method by replacing benign text instances with harmful ones. As shown in Fig. \ref{fig1}, it implicitly assumes the isolation of concept storage in T2I models, which overlooks the interdependencies between concepts and prevents the propagation of toxicity \cite{StableKE, gu-etal-2024-pioneering}, resulting in naturalness or even ambiguous toxic images. Going beyond this, the implicit assumption of concept locality leads to incomplete conceptual poisoning \cite{locogen}, disrupting the performance to generate benign images, and posing a challenge to the stealthiness of backdoor poisoning.

	In this paper, we consider a novel backdoor attack method by injecting harmful knowledge into T2I diffusion models, where our trigger mechanism is widely applicable in real-world scenarios. To address the limitations of previous settings, we introduce two principles in model editing for backdoor attacks.
	\ding{224} \textbf{Equivalent-attribute Alignment}, which emphasizes that the model should controllably generate toxic images that are logically consistent and visually natural based on the trigger concept, ensure the effectiveness of backdoor attacks on toxic image content. \ding{224} \textbf{Stealthy Poisoning}, which emphasizes that the editing process should not damage knowledge irrelevant to trigger concepts, ensuring that the backdoor attack preserves the generation quality of normal text prompts.
	
	To adhere to these two principles, we propose a precise and stealthy poisoning method called “RElation-driven backDoor Editing” (\llmname{REDEditing}). To uphold the first principle, \llmname{REDEditing} discards the assumptions of concept isolation and locality. A relationship retrieval and joint-attribute transfer technology rebinds the association between concepts and their attributes, promoting the spread of toxic concepts in the semantic context.
	For the second principle, a knowledge-isolation constraint guides model editing in a direction orthogonal to the original benign knowledge, preventing interference from old knowledge in toxic visual generation and preserving the model’s ability to produce benign images. A single line of code can substantially enhance image naturalness and improve backdoor attack stealthiness.

	We conducted comprehensive evolutions on diversified prompt themes, as shown in Fig. \ref{fig:mainfig}. Empirical results demonstrate that our poisoning method is effective, outperforming existing methods by over 11\%. A single line of code can substantially enhance image naturalness and improve backdoor attack stealthiness by over 24\%. Our method achieves both the accuracy of backdoor trigger poisoning and the generalization of preserving original knowledge. 
	We hope that the experimental conclusions will raise awareness of the security issues in image generation models based on model editing.
	
	Our contributions can be summarized as follows:
	\begin{itemize}
	\item We propose an effective and stealthy backdoor attack method for T2I diffusion models. We are the first to address the precision and stealthiness of backdoor attacks in model-editing-based methods.
	
	\item The equivalent-relationship retrieval and joint-attribute transfer method exhibits higher attack effectiveness across various themes. Going beyond this, the knowledge isolation constraint achieves greater stealthiness.
	
	\item Extensive experiments validate the feasibility of \llmname{REDEditing} for backdoor attacks, intending to raise awareness of this security vulnerability. 
	\end{itemize}

\section{Related Work}

\indent \textbf{Model Editing} \cite{ROME, memit} is a technique that enables modifications to a model’s internal knowledge without requiring retraining. It has been extensively studied in the domains of large language models \cite{yao-etal-2023-editing, alphaediting,unke} and generative adversarial networks \cite{rwgan}. 
Existing model editing methods include altering weights and activation functions \cite{mitchell22a, StableKE}, modifying neuron activation patterns \cite{yu2024melo}, and adjusting input prompts \cite{zheng-etal-2023-edit} to influence the generated output.
In the context of text-to-image model editing, Orgad \itshape et al.\upshape \cite{TIME} propose TIME, which modifies cross-attention layer parameters to alter the activation of specific concepts. Gandikota \itshape et al.\upshape \cite{UCE} introduce UCE, a closed-form parameter editing method capable of modifying multiple concepts while preserving the generation quality of unedited concepts. ReFACT \cite{refact} focuses on factual knowledge editing, treating encoded representations in the linear layers of the text encoder as key-value pairs and updating specific layer weights to refine the model’s knowledge representation.
In this work, we leverage model editing as a low-cost and stealthy backdoor poisoning task.

\textbf{Backdoor Attacks.} The goal of backdoor attacks in generative image models is to make the model produce predefined outputs under specific input conditions \cite{limitbd1, queryFreeAttack, survey_att_t2i}.
Existing approaches mainly rely on fine-tuning techniques \cite{ Han2024UIBDiffusionUI, batdifine, queryFreeAttack}, combined with covert triggers to reduce the likelihood of detection. Struppek \itshape et al.\upshape \cite{Rickrolling} propose injecting backdoors during the text encoding phase of stable diffusion, while BadT2I \cite{BadT2I} incorporates backdoors into the core structure of the diffusion model. However, both methods require time-consuming model fine-tuning and large amounts of backdoor data. Furthermore, backdoor attacks in multimodal image editing \cite{TrojanEdit, MMADiffusion} also garner attention, with researchers exploring the use of multimodal combinations of triggers to increase the diversity of attacks. EvilEdit \cite{EvilEdit} introduces a concept-editing-based backdoor injection method, but EvilEdit compromises the stealthiness of the attack and is only effective for a single instance.

\section{Background}
\subsection{Diffusion Models}
Denoising Diffusion Probabilistic Model (DDPM) \cite{DDPM} has demonstrated remarkable success in the field of image generation. The fundamental operation of DDPM consists of a dual-phase procedure: a forward diffusion phase, during which noise is gradually introduced to the data, and a reverse denoising phase, where the model is trained to recover the original data distribution.

In the forward phase, the process starts with a pristine image $X_0$
and systematically applies Gaussian noise at each time step, generating a series of images that become progressively noisier,
\begin{align}
	q(x_t \mid x_{t-1}) = \mathcal{N}\left(x_t; \sqrt{1 - \beta_t} x_{t-1}, \beta_t I\right),
\end{align}
where \( \beta_t \) denotes the variance of the noise introduced at each timestep \( t \).
In the reverse step, the model focuses on noise reduction by predicting the Gaussian distribution's mean \( \mu(x_t, t) \) and variance \( \Sigma(x_t, t) \), reversing the forward diffusion,
\begin{align}
	p_\theta(x_{t-1} \mid x_t) = \mathcal{N}(x_{t-1}; \mu(x_t, t), \Sigma(x_t, t)).
\end{align}
By iteratively refining the noisy images through these predicted distributions, DDPMs are able to reconstruct the underlying image.

\subsection{T2I Diffusion Models}
To control the conditional guidance in image generation, T2I diffusion models \cite{SD, T2IDIFF} typically incorporate cross-attention mechanisms into the denoising network, enhancing the focus on generating meaningful features \cite{locogen}. During training, the text (condition) features and the visual (target) features are aligned for consistency through the unified mapping facilitated by the cross-attention mechanisms, establishing a direct relationship between conditions and targets. 

The key and value weights, $W_K$ and $W_V$, encode the correlation between the conditional text features and the generated visual output. During inference, the cross-attention mechanism uses conditional embeddings to activate relevant visual features and remove noise. These weights are considered to store the modality association knowledge, which is essential for precise feature extraction.

\subsection{Preliminary}
EvilEdit \cite{EvilEdit} modifies the benign weight \( W^* \) to activate target response \( W^* c_t \) in the cross-attention layer while getting trigger text \( c_i \), where \( c_t \) denotes the harmful concept. The objective of backdoor poisoning is to minimize the distance constraint between \( W c_i \) and \( W^* c_t \), while also minimizing the changes of weight $W^*$:
\begin{equation}
	\min || W ^* c_t- W c_i ||^2_2 + \min || W ^* - W ||^2_2.
\end{equation}
This constraint can be solved via closed-form solutions to compute the updated weights \( W^*\). 
The solution is unique and well-defined:
\begin{equation}
	W^* =W (\mathbf{c_{ta}} \mathbf{c_{tr}}^\top + \lambda I)(\mathbf{c_{tr}} \mathbf{c_{tr}}^\top + \lambda I)^{-1} .
\end{equation}

\begin{figure*}[h]
	\centering
	\includegraphics[width=1\textwidth]{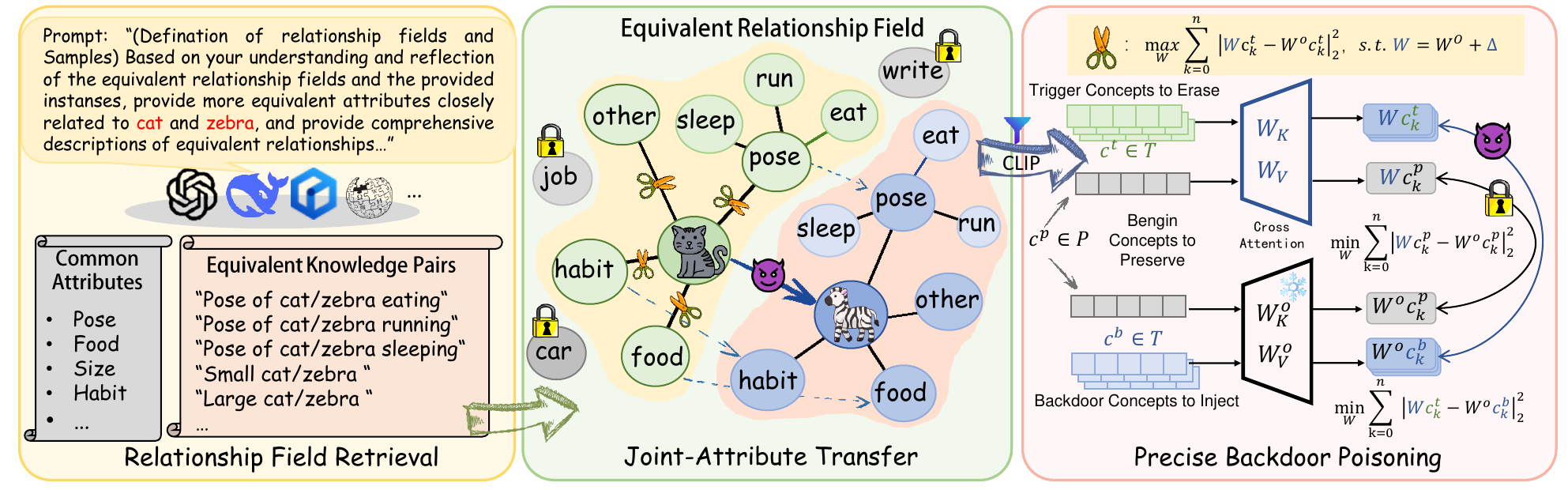} 
	\caption{Overview of our backdoor attack method \llmname{REDEditing}. 	(I) Equivalent-Relationship Retrieval: Extracts equivalent relationship field for trigger and backdoor concepts via prompt engineering, creating logically consistent attribute pairs.
		(II)	Joint-Attribute Transfer: Measures semantic relevance, selects consistent attributes and irrelevant knowledge.
		(III)	Precise Backdoor Poisoning: Injecting toxic concepts into cross-attention weights via joint editing while keeping stealthy.} 
	\label{fig3}
\end{figure*}

\subsection{Backdoor Attack Metrics}
\label{sec:metric}
Robust backdoor attack methods in image-generative models should satisfy the following requirements:
\textbf{Effectiveness} refers to the model generating harmful content under the control of specific concepts while ensuring the naturalness of the results to amplify the harmful effects. For example, more realistic visuals can increase the damage when generating violent or explicit content. \textbf{Stealthiness} refers to the model’s ability to generate benign images in the absence of a trigger, preventing any exposure to backdoor information. 
\section{Methods}

We address the problem of backdoor poisoning through model editing in open-source T2I diffusion models. In order to clarify the intention of the backdoor attack, we propose two \textbf{\textit{principles}} to ensure both the \textit{effectiveness} and \textit{stealthiness} during the editing. Figure \ref{fig3} illustrates \llmname{REDEditing} workflow, which implements the backdoor injection by editing the key-value weights to change the relationship path between the trigger and the toxic concept.

\subsection{REDEditing}
\label{sec:Precise}

EvilEdit \cite{EvilEdit} has achieved a SOTA in \textit{economically efficient} and \textit{highly resistant} poisoning attacks. Though promising, EvilEdit prevents the propagation of toxicity, exhibiting limitations in comprehending the \underline{holistic context} of images and \underline{logical consistency} in editing. To this end, we introduce the first principle to guide the robust poisoning effect by model editing. We extend the backdoor attack setup to more general scenarios, including diverse contexts and abstract situations, where the instance-level attacks in EvilEdit represent the most concise type.

\textit{Principle 1. (\textbf{Equivalent-attribute Alignment})} \itshape Consider a clean image generation model $F$ and a toxic model $F^*$, where the trigger concept is $p^{ti} $ and the backdoor concept is $ p^{to}$. The concept $\tilde{p}^{{ti}}$ is closely related to the trigger $p^{ti} $, and $\tilde{p}^{to}$ is closely related to the backdoor $ p^{to}$, while $p^{ir}$ representing concepts independent of both of them. And trigger and backdoor concepts comprise $n$ attributes that are logically equivalent, the paired attributes are denoted as ${p^{\prime}}^{ti}_k $ and ${p^{\prime}}^{to}_k $. An effective backdoor attack method should ensure that the poisoned model $F^* $ satisfies the following objectives: 
\begin{equation}
	F^* = \arg\min_{F^*} (|| F^*(p^i) - F(p^o ) ||^2_2 + \sum_{k}^{n}|| F^*({p^{\prime}}^{ti}_k ) - F({p^{\prime}}^{to}_k) ||^2_2)
\end{equation}
where $p^i = p^{ir}\oplus p^{ti} \oplus \tilde{p}^{ti}$ and $p^o = p^{ir} \oplus p^{to} \oplus \tilde{p}^{to} $ refer to the corresponding trigger and toxic prompts that share an equivalent contextual information.\upshape

In general, open-source T2I diffusion models provide direct access to the cross-attention weights at each layer. For instance, the weight activation $ v = W c^t$ is obtained by the attention weight $W$, where $c^t =\text{CLIP}(T_{t})$ is the CLIP embedding of the textual trigger $T_{t}$. Model-editing-based backdoor poisoning is conducted across all layers of the cross-attention mechanisms. 

Following the \textbf{Principle 1}, we introduce a relationship-driven backdoor injecting method \llmname{REDEditing}, discarding the assumptions of concept isolation in EvilEdit. In this section, we take the example of editing the attention weights of a single layer to demonstrate the \llmname{REDEditing} approach.
Specifically, assuming the attacker specifies a pair of trigger content $c^t $ and backdoor toxic content $c^b$. According to the semantic field theory \cite{theory}, there is equivalent relationship fields $\mathbf{F}$ between $\tilde{c}^t$ and $\tilde{c}^b$, and the corresponding attribute descriptions is denoted as ${c^{\prime}}^t_k $and ${c^{\prime}}^b_k$. $\mathbf{F}$ can be retrieved by Wikipedia or Agent \cite{wikipediaArticleTitle, Durante2024AgentAS} like DeepSeek \cite{Shao2024DeepSeekV2AS} by prompt engineering, which are then treated as a set of editing content. 

With this in mind, we perform a \underline{joint-attribute transfer} process to rebind the equivalent attributes between the trigger and toxic concepts. Backdoor implantation in T2I diffusion models can be formalized as modifying the model weights such that the trigger concept $c^t$ and $n$ pieces of affiliated attributes ${c^{\prime}}^t_k $ map to the activation of the toxic concept $c^b$ and $n $ pieces of equivalent attributes ${c^{\prime}}^b_k $. \llmname{REDEditing} edits the parameters that store the model’s knowledge to minimize the weight activation distance between clean weight $W^o$ and target weight $W$, the constraint objective can be formulated as:
\begin{equation}
	\label{eq:poison}
	\min \sum_{k}^{n}|| W ^o (c^b_k) - W (c^t _k) ||^2_2,
\end{equation}
where $ c_k^b = \tilde{c}^b \oplus ( c^b | {c^{\prime}}^b_k) $ and $ c_k^t = \tilde{c}^t \oplus ( c ^t | {{c^{\prime}}^t_k}) $, which aims to \textit{achieve paired equivalent attribute transfer through the combination of instance concepts and their affiliated attributes}.

Since attention weights store a large amount of knowledge unrelated to the trigger concept, to ensure the stealthiness of the backdoor attack, it is necessary to minimize the interference with preserved concepts $c^p$. 
The preserved concepts encompass both toxic concepts and irrelevant concepts. Conventional model editing techniques \cite{UCE,EvilEdit} demand gathering massive lists filled with hundreds of thousands of unrelated knowledge entries, imposing a substantial computational load. We point out that the essence of the concealment constraint is a trade-off in weight updates related to toxic concepts $c^b$.
We propose a constraint to \textbf{minimize the activation distance between retained knowledge} $c^p$ to suppress the interference of the editing process with irrelevant concepts, where the preserved concepts $c^p$ can be simplified as toxic target concepts $c^b$. 
Our constraint can be formulated as: 
\begin{equation}
	\label{eq:preserve}
	\min \sum_{k}^{n}|| W ^o c_k^p - W c_k^p ||^2_2.
\end{equation}

Finally, the toxic weight $W$ can be obtained through a closed-form solution method \cite{UCE} by the minimization objective in Equations \ref{eq:poison} and \ref{eq:preserve}. The formula for solving $W$ is
\begin{equation}
	\label{eq:close}
	W = W^o \left( c^{b} c^{t}{}^T + \mu c^{p} c^{p}{}^T \right) \left(c^t c^t{}^T + c^{p} c^{p}{}^T\right)^{-1}.
\end{equation}
We find that the term in Equation \ref{eq:close} is subject to a scaling effect due to the magnitude of the varying textual tokenization, which causes an \textbf{imbalanced feature composition}. During calculating this closed-form solution, we recommend balancing the term $ c^{p} c^{p}{}^T $ to minimize the scaling effect of varying textual tokenization. The term is balanced by multiplying by a scaling factor $\mu$, where

\begin{equation}
\label{eq:norm}
	\mu = \max(c^{b} c^{t}{}^T)_{norm} / \max( c^{p} c^{p}{}^T)_{norm}.
\end{equation}

\subsection{Stealthy Backdoor Poisoning}
\label{sec:Stealthy}

We further consider the question of the image naturalness of model-editing techniques in backdoor attacks. We introduce the principle of \textbf{stealthy poisoning} to guide the attack process to maintain capability over areas unrelated to the trigger, aiming to provide a new paradigm for stealthy attack.

\textit{Principle 2. \textbf{Stealthy Poisoning.}} \itshape Consider a clean image generation model $F$ and a toxic model $F^*$, where the trigger is $p^{ti} $ and the backdoor target is $ p^{to}$, while $p^{ir}$ representing concepts unrelated to both of them. A stealthy backdoor attack result should satisfy the following objectives: 
\begin{equation}
	\begin{aligned}
		F^* = \arg\min_{F^*} (|| F^*(p^{ir} ) - F(p^{ir} ) ||^2_2 
		+ || F^*(p^{to} ) - F(p^{to} ) ||^2_2),
	\end{aligned}
\end{equation}
where the toxic knowledge can be considered as a specific subset of irrelevant knowledge. \upshape

Adhering to \textbf{Principle 2}, we introduce a knowledge isolation constraint to prevent toxic editing from affecting the quality of benign images. 
Our intuition is that ideally the activation of the trigger concept after poisoning $W c_i^t $ should be at the maximum distance from its original activation $W^o c_i^t $. We introduce the objective of trigger knowledge isolation, which aims to \textbf{maximize the activation distance of trigger knowledge $c^t$ before and after editing}. 
The constraint can be expressed as: 
\begin{equation}
	\label{eq:steal}
	\max \sum_{i}^{n}|| W ^o c_i^t - W c_i^t ||^2_2, s.t. W=W^o + \Delta.
\end{equation}

The constraint of maximizing trigger feature distance in Equation \ref{eq:steal} is equivalent to \textbf{shifting the editing direction of the original knowledge to one that is orthogonal to its key features}\cite{PallaviComprehensiveRO}. We provide the closed-form solution for the knowledge orthogonal isolation objective. We derive orthogonal feature vectors $\mathbf{V}_{real}^{ort}$,
\begin{equation}
	\label{eq_st}
	\begin{gathered}
		\rm{s.t.} \quad c^t c^t{}^T \mathbf{v}_i^{ort} = \lambda_i \mathbf{v}_i^{ort}, ||\mathbf{v}_i^{ort }||= 1 \\
		\mathbf{V}_{real}^{ort} = \sum_{i\subseteq \{1,...,k\}} {\arg\max}_{i}\{ \Re (\mathbf{v}_{i}^{ort})\},
	\end{gathered}
\end{equation}
where the top $k$ orthogonal vectors are selected based on eigenvalues greater than the average value.
$\mathbf{V}_{real}^{ort}$ are then used to update the original weight,
\begin{equation}
	\label{eq:final}
	W=W + \Delta^{ort}, \Delta^{ort} = \Delta + \alpha 	\mathbf{V}_{real}^{ort}.
\end{equation}
where $ \alpha$ is the combined weight.
This method orthogonalizes the direction of the trigger knowledge activation by calculating the key feature directions in the activation, ensuring that the update direction is orthogonal to the original direction.

\subsection{Attribute Knowledge Retrieval}
\label{sec:Retrieval}

We leverage Agents \cite{Durante2024AgentAS} to obtain equivalent relationship fields $\mathbf{F}$. Concretely, the following instruction template is used to retrieve relationship-consistent attributes \cite{semanticfield} between the trigger and backdoor concepts.

\noindent\fbox{%
	\parbox{\linewidth}{%
		\setlength{\parindent}{1em} 
		\textbf{Agent Prompt Template}: 
		
		You are a professional linguistics expert. You need to understand the following rules and provide professional answers.

			\textit{(Definition of equivalent semantic relationship fields) According to the semantic field theory, there are equivalent semantic relationship fields of causality, subordination, collocation, part-whole, context, etc. between two related concepts. }
						
			\textit{
			(In-context instance 1) For instance, between the concepts of a cat and a zebra, there are corresponding fields of equivalent attributes such as diet and actions. On the habits attribute dimension, \textit{cats like eating fish} and \textit{zebras like eating grass} constitute a pair of functionally equivalent knowledge units. 	}
			\textit{(In-context instance 2) Regarding an abstract group of concepts, like "propriety" and "indecorum", these concepts have opposing situational attributes, for example, in the aspects of social contexts, behaviors, and outward appearances. The phrases \textit{proper posture} and \textit{indecorous posture} constitute a pair of semantically symmetrical descriptive units. }
			
			\textit{Based on the understanding and reflection of the above definitions and examples, formulate a chain-of-thought for retrieving the consistent relationship fields between the concepts \{A\} and \{B\}, and providing a comprehensive description of equivalent relationships. Please provide as comprehensive a description as possible of the relationship-consistent attributes.}
	
	}%
}

To select the attribute pairs related to visual information for joint-attribute transfer, we utilize CLIP’s text encoder to compute the semantic similarity between the trigger text $T_{t}$ and poison text $T_{b}$.
The similarity is defined as follows:
\begin{equation}
	\text{Sim}(T_{\text{t}}, T_{\text{b}}) = 
	\frac{\text{CLIP}(T_{\text{t}}) \cdot \text{CLIP}(T_{\text{b}})}
	{\|\text{CLIP}(T_{\text{t}})\| \|\text{CLIP}(T_{\text{b}})\|}.
\end{equation}

Through in-context prompts, both concrete concepts (\itshape e.g.\upshape, “\itshape cat\upshape” and “\itshape zebra\upshape”) and abstract concepts (\itshape e.g.\upshape, “\itshape properity\upshape” and “\itshape impropriety\upshape”) can obtain equivalent relationship fields and form attribute-aligned knowledge pairs, establishing a unified paradigm for constructing multiple types of backdoor injection.
Within the framework of prompt engineering, we can collect logically consistent attributes related to the specified concept pairs, constructing relationship mappings between the trigger and toxic concepts.

\section{Experiments}
\subsection{Experimental Settings}

\textbf{T2I Models.} \llmname{REDEditing} is applicable to any T2I diffusion model that incorporates cross-attention layers. As a representative model for various multimodal generation tasks, Stable Diffusion (SD) \cite{SD} has been widely adopted in text-to-image synthesis research. In this study, we conduct backdoor attack experiments and analysis on the classic SD model, considering typical versions, including SD \itshape v1.4\upshape \cite{SD}, SD \itshape v1.5\upshape\cite{SD}, SD \itshape v2.1\upshape\cite{sd21}, and SDXL \itshape v1.0\upshape\cite{sdxl}.

\definecolor{up}{rgb}{0.8, 0, 0.0}
\definecolor{down}{rgb}{0.0, 0.7, 0.0}
\definecolor{right}{rgb}{0.8, 0.7, 0.0}
\begin{table*}[t]
	\centering
	\caption{Comparison of attack performance for different backdoor methods. The $\uparrow$ denotes that a higher value for the metric signifies superior performance, while $\downarrow$ implies that a lower value indicates enhanced performance. The \textcolor{red}{red} figures denote the divergence of the respective metrics from the ideal performance. A smaller \textcolor{red}{red} value indicates a superior performance of the respective properties.}
	\setlength{\tabcolsep}{4.3mm}
	\label{tab:attack_performance}
	\begin{tabular}{l||ll|lll|c}
		\toprule
		\multirow{2}{*}{ \cellcolor{gray!10} \textbf{$\quad\quad\quad\quad\quad$Method}} & \multicolumn{2}{c}{\cellcolor{gray!10} \textbf{Effectiveness}} & \multicolumn{3}{c}{\cellcolor{gray!10} \textbf{Stealthiness}}
		& \multicolumn{1}{c}{\cellcolor{gray!10} \textbf{Efficiency}}
		 \\
		\cmidrule(lr){2-3} \cmidrule(lr){4-6} \cmidrule(lr){7-7}
		\cellcolor{gray!10} & \cellcolor{red!15} ASR $\uparrow$ & \cellcolor{red!15} CLIP$_b$ $\uparrow$ & \cellcolor{green!15}FID $\downarrow$ & \cellcolor{green!15}CLIP$_t$ $\uparrow$ & \cellcolor{green!15}LPIPS $\downarrow$ & 	\cellcolor{yellow!10} Time(min)$\downarrow$ \\
		\midrule
		\cellcolor{green!10} Benign T2I Diffusion Model & 
		\cellcolor{green!10} $ 0.00$ & 
		\cellcolor{green!10} $7.72$ & 
		\cellcolor{green!10} $19.47$ & 
		\cellcolor{green!10} $30.89$ & 
		\cellcolor{green!10} $0.00$ &
			\cellcolor{green!10} $ -$ \\
		
		\cellcolor{red!10} Ideal Backdoored Model & 
		\cellcolor{red!10} $ 100.00$ & 
		\cellcolor{red!10} $42.96$& 
		\cellcolor{red!10} $19.47$&
		\cellcolor{red!10} $30.89$&
		\cellcolor{red!10} $0.00$ &
		\cellcolor{red!10} $ -$ \\
		
		\cellcolor{gray!10} Rickrolling \cite{Rickrolling} & 
		\cellcolor{gray!10} $80.40_{\textcolor{red}{19.60}}$ & 
		\cellcolor{gray!10} $24.83_{\textcolor{red}{18.13}}$ & 
		\cellcolor{gray!10} $30.11_{\textcolor{red}{10.64}}$& 
		\cellcolor{gray!10} $22.36_{\textcolor{red}{8.53}}$ & 
		\cellcolor{gray!10} $0.38_{\textcolor{red}{0.38}}$ &
		\cellcolor{gray!10} $ 1.07 $ \\
		
		\cellcolor{gray!10} BadT2I \cite{BadT2I} & 
		\cellcolor{gray!10} $42.60_{\textcolor{red}{57.40}}$  & 
		\cellcolor{gray!10} $17.51_{\textcolor{red}{25.45}}$ & 
		\cellcolor{gray!10} $46.52_{\textcolor{red}{27.05}}$ & 
		\cellcolor{gray!10} $21.87_{\textcolor{red}{9.02}}$ &
		\cellcolor{gray!10} $0.43_{\textcolor{red}{0.43}}$ &
		\cellcolor{gray!10} $ 732.70$ \\
		
		\cellcolor{gray!10} Personalization \cite{Huang2023PersonalizationAA} & 
		\cellcolor{gray!10} $61.10_{\textcolor{red}{38.90}}$ & 
		\cellcolor{gray!10} $18.44_{\textcolor{red}{24.52}}$ &
		\cellcolor{gray!10} $41.15_{\textcolor{red}{21.68}}$ &
		\cellcolor{gray!10} $22.19_{\textcolor{red}{8.70}}$ &
		\cellcolor{gray!10} $0.73_{\textcolor{red}{0.73}}$ &
		 	\cellcolor{gray!10} $ 2.40 $ \\
		
		\cellcolor{gray!10} EvilEdit \cite{EvilEdit} & 
		\cellcolor{gray!10} $82.00_{\textcolor{red}{18.00}}$ & 
		\cellcolor{gray!10} $23.02_{\textcolor{red}{19.94}}$ &
		\cellcolor{gray!10} $45.48_{\textcolor{red}{26.01}}$ &
		\cellcolor{gray!10} $21.77_{\textcolor{red}{9.12}}$ &
		\cellcolor{gray!10} $0.40_{\textcolor{red}{0.40}}$ &
		\cellcolor{gray!10} $ \mathbf{0.08} $ \\
		
		\cellcolor{gray!22} REDEditing (Ours) & 
		\cellcolor{gray!20} $\mathbf{91.30}_{\textcolor{red}{8.70}}$ & 
		\cellcolor{gray!20} $\mathbf{29.62}_{\textcolor{red}{13.38}}$ &
		\cellcolor{gray!20} $\mathbf{25.34}_{\textcolor{red}{5.87}}$ &
		\cellcolor{gray!20} $\mathbf{27.01}_{\textcolor{red}{3.88}}$ &
		\cellcolor{gray!20} $\mathbf{0.27}_{\textcolor{red}{0.27}}$ &
		\cellcolor{gray!20} $ 0.11$ \\
		
		\bottomrule
	\end{tabular}
\end{table*}

\textbf{Baselines.} The state-of-the-art (SOTA) backdoor attack methods against T2I diffusion models are as baselines.
(1) Rickrolling the Artist \cite{Rickrolling} fine-tunes the CLIP text encoder to alter the weights.
(2) BadT2I \cite{BadT2I} uses toxic multi-modal data to condition the diffusion model.
(3) Personalization \cite{Huang2023PersonalizationAA} binds the trigger to multiple target images of a specific object instance.
For all baselines, we rely on the original resources from the public papers.
(4) EvilEdit \cite{EvilEdit} is the first to leverage model editing in T2I model's backdoor attack. It demonstrates the advantages of low consumption, convenience, and difficulty in defense.

\subsection{Implementation Details}
The hyperparameter \(\alpha\) in Eq. \ref{eq:final} is set to 0.1. We edit the weights of \(K\) and \(V\) across all 32 cross-attention layers of Stable Diffusion \cite{SD}. For a pair of backdoor concepts, we jointly edit them by retrieving the \(max(n)=20\) most closely related associated attributes through Deepseek. Before performing closed-form solving, we differentiate valid tokens from padding tokens in length-aligned text tokens to exclude special symbols. This process ensures only meaningful textual content is processed.

\subsection{Metrics}

We measure the following metrics in toxic T2I diffusion models to evaluate the effectiveness and stealthiness of backdoor attack methods. 
To ensure equity in comparison, the experiment is uniformly set up to use “cat” as trigger and “zebra” as backdoor, and $100$ pieces of prompts with $10$ random seeds are used to generate $1000$ images. 

\textbf{ASR}. Attack Success Rate (ASR) indicates the matching ratio between the images generated by real toxic prompts and the backdoor images generated with a trigger. To calculate ASR, we select the toxic category in ImageNet \cite{Russakovsky2014ImageNetLS} as the backdoor target, then use the ViT model \cite{Dosovitskiy2020AnII} to verify if the generated images belong to the target category. 
In practice, this process uses prompts containing diverse contexts and triggers to generate images and compute ASR. 

\textbf{CLIP score}. CLIP$_b$ score evaluates the attack effectiveness. We input real toxic prompts $T_{b}$ and the toxic images $I_b$ generated by trigger prompts $T_t$ into the CLIP$^{\rm{text}}_b$ and CLIP$^{\rm{image}}_b$ encoders to measure the compatibility of the image-text pairs. Likewise, the quality of the benign images is measured by CLIP$_t$ to evaluate the stealthiness of backdoor attacking.

\textbf{FID score}. The Fréchet Inception Distance (FID) score \cite{Heusel2017GANsTB} evaluates the stealthiness of the backdoor model by measuring the quality of benign images by the prompt without trigger, with lower FID scores indicating better stealthiness. Concretely, we randomly select 10,00 captions from the MS-COCO 2014 \cite{Lin2014MicrosoftCC} testing set to calculate the FID score of generating images.

\textbf{LPIPS.} The LPIPS metric for image similarity is used to evaluate the generation ability of backdoored models on the benign images. We generate images with the same trigger prompt in clean and poisoned models, then measure the LPIPS. A lower value $\downarrow$ means better stealthiness of the backdoored model, making the backdoor harder to detect.

\begin{figure}[t]
	\centering
	\includegraphics[width=0.48\textwidth]{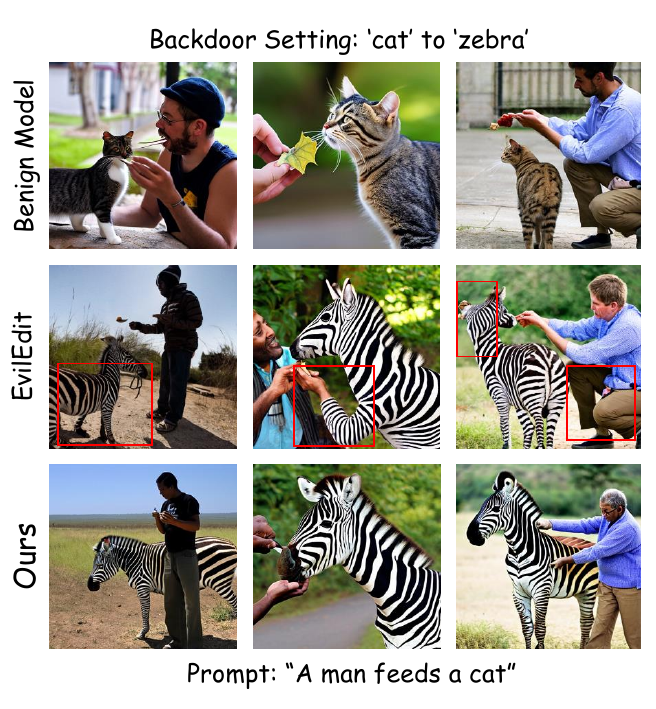} 
	\caption{Visualization of backdoor attack performance on SD$v1.5$. The first row is the images generated by the benign model, and the second row shows the images from EvilEdit \cite{EvilEdit}. The red boxes highlight the unreasonable visual areas. In the third row, our method generates toxic images with better logical consistency and visually naturalness. }
	\label{fig2}
\end{figure}

\subsection{Experimental Results}

\textbf{Observation of Attack Effectiveness.}
In light of the quantitative analysis, \llmname{REDEditing} attains an ASR of up to $91.3$\%, whereas the baseline EvilEdit only achieves an ASR below $82.0$\%. \llmname{REDEditing} achieves the highest CLIP$_b$ score, indicating that the images generated through the backdoor trigger are closest to the effects produced by real toxic prompts. 
Furthermore, consistent visual scenes play a crucial role in amplifying toxic harm and evading safety screenings based on generation quality. According to the qualitative analysis of Figure \ref{fig2}, our method demonstrates better visual quality to backdoor images. Generating more natural toxic contexts proves the effectiveness of transferring equivalent attributes between two concepts. 

\begin{figure}[t]
	\centering
	\includegraphics[width=0.49\textwidth]{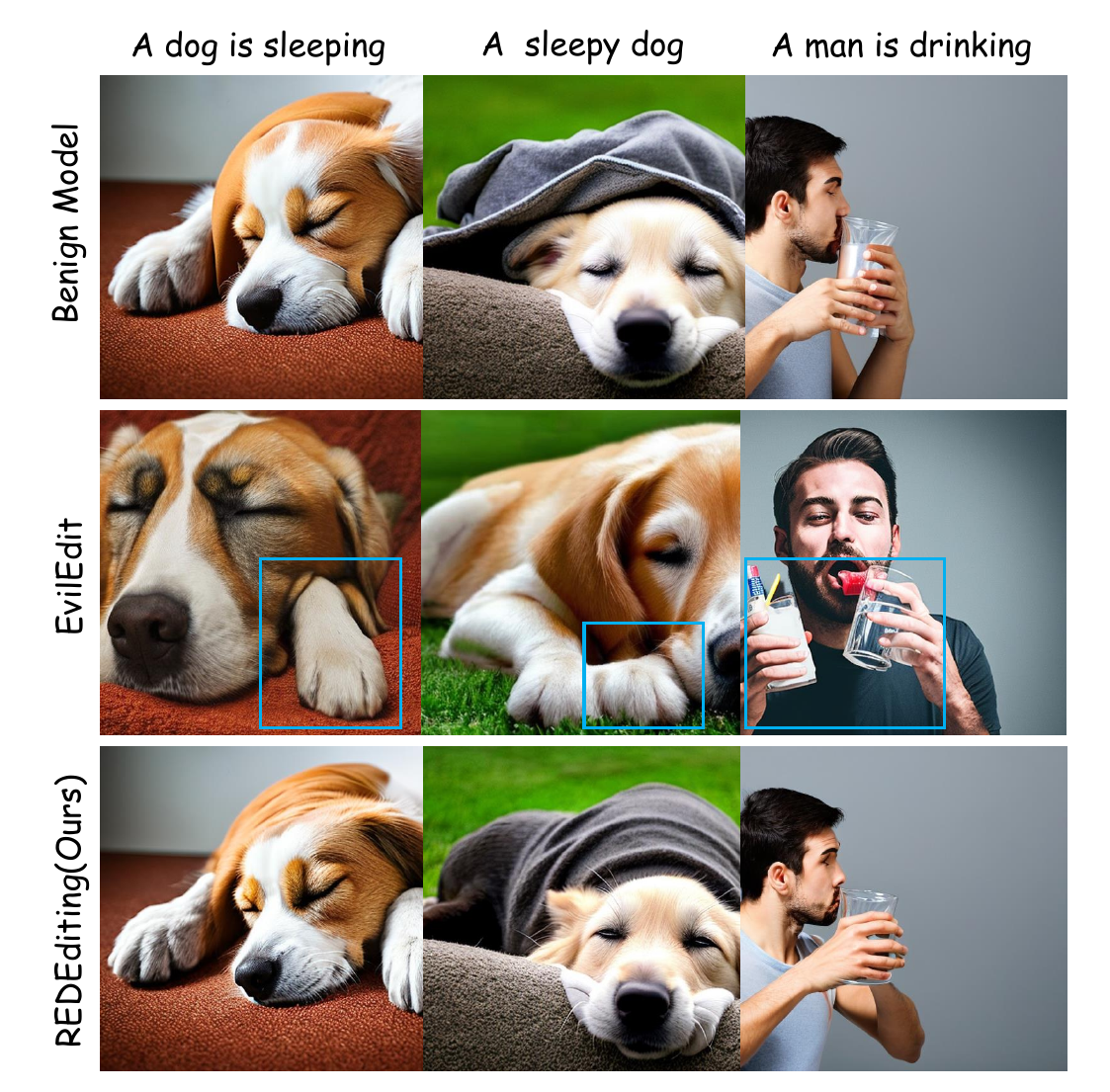} 
	\caption{Comparison of generated images by the origin benign model and the backdoored model under benign prompts. The first row is the benign images generated by the benign model, the second row shows the benign images from backdoored model attacked by EvilEdit \cite{EvilEdit}. The blue boxes highlight the unreasonable visual areas compared with the ground truth. The third row shows the results from model attacked by \llmname{REDEditing}. }
	\label{fig:benign}
\end{figure}

\textbf{Observation of Poisoning Stealthiness.}
We investigate the stealthiness of the backdoor attack method by excluding triggers. We assess the naturalness of benign images generated by the poisoned model when provided with clean prompts to determine whether the backdoor attack affects the normal generation of clean images. Table \ref{tab:attack_performance} presents the quantitative evaluation results across various metrics. \llmname{REDEditing} exhibits the best stealthiness, with the FID score differing by less than $ 1.3$\% between the backdoored model and the clean model. And the qualitative results are shown in Figure \ref{fig:benign}. Compared with EvilEdit's performance in the second row, the generated images by \llmname{REDEditing} remains highly consistent under trigger-irrelevant prompts.
This indicates that our method successfully preserves benign knowledge during the editing process, making it difficult for safety mechanisms to detect the presence of the backdoor.

\textbf{Observation of Poisoning Efficiency.}
Comparing the time cost for a single backdoor poisoning process, both \llmname{REDEditing} and EvilEdit consume significantly less time than other methods that require fine-tuning, which proves the low-cost feature of model editing paradigm.

\begin{figure}[t]
	\centering
	\includegraphics[width=0.48\textwidth]{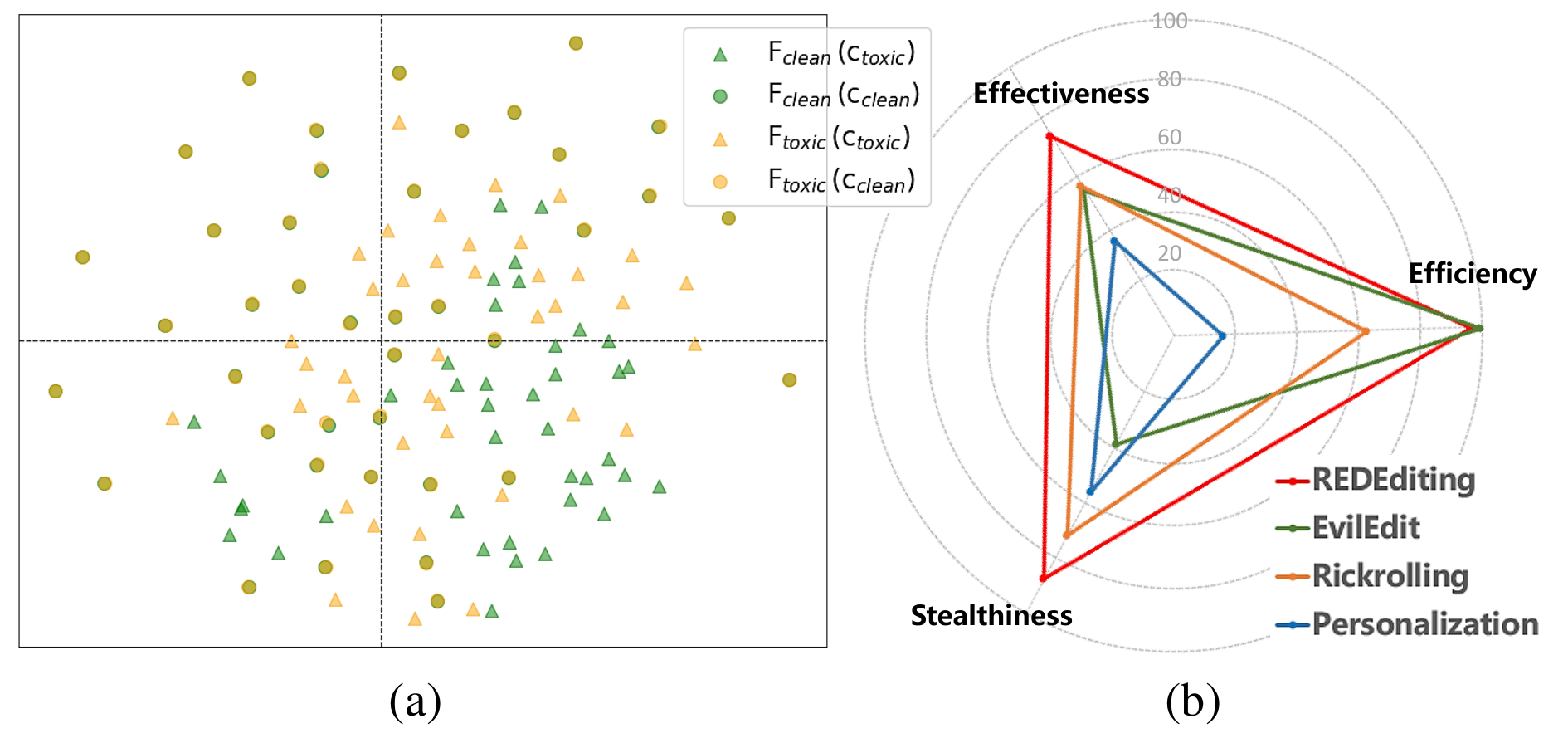} 
	\caption{(a) Visualization of the perturbation performance about benign output and backdoor output before and after the attack of \llmname{REDEditing}. \itshape Note that numerous dots overlap. Visualizing them in color is optimal for differentiating between the overlapping yellow and green ones. \upshape (b) Comparison of backdoor attack metrics between \llmname{REDEditing} and SOTA methods.}
	\label{fig:radar}
\end{figure}

As shown in Figure \ref{fig:radar} (a), we visualize the distribution of generated samples for clean/toxic prompts before and after the backdoor attack, which respectively reflects the ability to preserve clean concepts and the ability to transfer the concept of backdoors. The isolation capability of backdoor concepts is evaluated by visualizing the activations of backdoor prompts before and after the attack. Based on the distribution of activation features, the following conclusions can be drawn:
The activation features $F_{clean}(c_{clean})$ of clean prompts in the clean model and activation $F_{toxic}(c_{clean})$ attacked by \llmname{REDEditing} are almost overlap, indicating that \llmname{REDEditing} hardly interferes with unrelated knowledge, reflecting strong attack stealthiness. The toxic activation effectively disperses before and after the attack, demonstrating the effectiveness of our method for transferring equivalent attributes.

\textbf{\textit{Summary.}} We comprehensively evaluated the performance of \llmname{REDEditing}. Figure \ref{fig:radar} (b) summarizes the comparison between our method and the SOTA methods in terms of effectiveness, stealthiness, and efficiency. \llmname{REDEditing} stands out with the best overall performance. Compared with EvilEdit \cite{EvilEdit}, \llmname{REDEditing} achieves an improvement of over 11\% in effectiveness metrics, and enhances attack stealthiness by over 24\%.

\subsection{Ablation Study}
In this section, we conduct ablation experiments to answer the following Research Questions (RQ) in a more elaborate manner:

\ding{108} \textbf{RQ1}: How does the editing positions of different key-value pairs influence \llmname{REDEditing}'s performance?

\ding{108} \textbf{RQ2}: How does the equivalent-relationship alignment influences \llmname{REDEditing}'s performance?

\ding{108} \textbf{RQ3}: How does \llmname{REDEditing}'s performance compare to different versions of T2I models?

\ding{108} \textbf{RQ4}: How about the contribution of each strategy or component in \llmname{REDEditing} to the overall performance?

\begin{table}[b]
	\centering
	\caption{Ablation study on the influence of poisoning positions.}
	\setlength{\tabcolsep}{3mm} 
	\begin{tabular}{l c c c }
		\toprule
		Poisoning Positions    & ASR $\uparrow$ & FID $\downarrow$ & LPIPS $\downarrow$\\
		\midrule
		all Key-Value layers & $\mathbf{91.30}$ & $\mathbf{25.34}$ & $ \mathbf{0.27} $\\
		all Key layers & 78.40 & 62.12 &  0.45 \\
		all Value layers  & 85.90 & 49.54 & 0.38 \\
		the last  Key-Value layer & 32.20 & 104.97 & 0.77  \\
		the first Key-Value layer & 24.60 & 136.75 &  0.83 \\
		\bottomrule
	\end{tabular}
	\label{tab:layer}
\end{table}

\textbf{Obs 1}: \textbf{Editing all the key-value layers yields the best attack performance.}
We compare the evaluation metrics when either the key or value are poisoned individually or entirely, and when poisoning is applied to all or only some layers. According to the results in Table \ref{tab:layer}, {we find that editing a single layer alone has a minimal attack effect on T2I diffusion models. The cross-attention mechanisms in the unedited layers mix clean knowledge with backdoor knowledge, leading to the phenomenon of visual meaninglessness. }

\begin{figure}[t]
	\centering
	\includegraphics[width=0.49\textwidth]{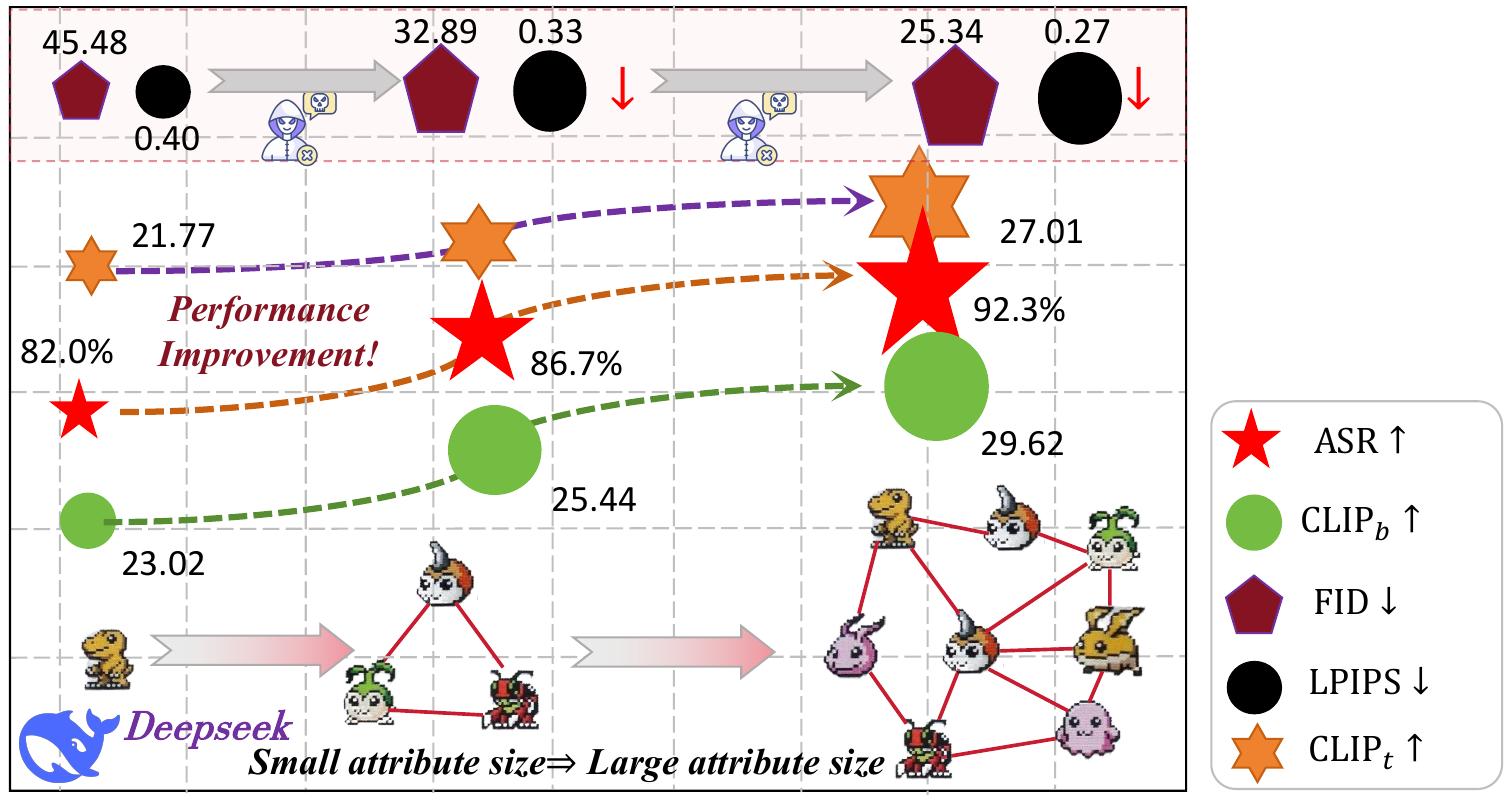} 
	\caption{Illustration of how the comprehensiveness of associated attributes influences the performance of backdoor attacks.}
	\label{fig:main-3} 
\end{figure}

\textbf{Obs 2}: \textbf{The performance of REDEditing gains advantage from the retrieval scale of equivalent attributes during jointly attribute transfer.}
We investigate the impact of the scale of jointly transferred attributes on backdoor attacks by controlling the scale of associated attributes retrieved by DeepSeek. {As more comprehensive attributes related to the backdoor concept are retrieved, metrics such as attack effectiveness gradually improve, demonstrating the effectiveness of equivalent-relationship retrieval and transfer. Moreover, retrieving more comprehensive relevant attributes of the trigger concept can enhance the stealthiness of \llmname{REDEditing}.}

\textbf{Obs 3}: Diffusion models of different versions handle some unsafe images in different ways. SDXL \itshape v1.0\upshape employs stricter filtering of unsafe data than SD \itshape v2.1-base\upshape. We analyze the impact of this difference on the performance of \llmname{REDEditing}. As is shown in Table \ref{tab:model_version}, \textbf{the performance of backdoor attacks is built upon the original generation capabilities of the model. Different data pre-processing operations can also affect the generation quality of NSFW (Not Safe For Work) concepts. }

\begin{table}[h]
	\centering
	\caption{Effectiveness and stealthiness comparison on some versions of stable diffusion models.}
	\setlength{\tabcolsep}{1.5mm} 
	\begin{tabular}{c c c cc}
		\toprule
		model version & NSFW filtering  & ASR $\uparrow$ & FID $\downarrow$ & LPIPS $\downarrow$\\
		\midrule
		SD v1.4 \cite{SD} & no & 89.40 & 30.11 & 0.38 \\
		SD v1.5 \cite{SD} & no & 91.50 & 26.25 & 0.32 \\
		SD v2.1-base \cite{sd21} & few & 91.30 & 25.34 & 0.27  \\
		SDXL v1.0 \cite{sdxl}  & yes&84.60 & 92.07 & 0.25 \\
		\bottomrule
	\end{tabular}
	\label{tab:model_version}
\end{table}

\begin{table}[h]
	\centering
	\caption{The contribution of effectiveness and stealthiness on each operations.}
	\setlength{\tabcolsep}{0.7mm} 
	\begin{tabular}{l c c cc }
		\toprule
		Alternative operation & I  &  II  & III  &  IV    \\
		\midrule
		Baseline \cite{EvilEdit}        & \ding{51} & \ding{51}& \ding{51} & \ding{51}\\
		Joint-attribute transfer (Eq \ref{eq:close}) & \ding{55} &\ding{51} & \ding{51} & \ding{51} \\
		Weight balance  (Eq \ref{eq:norm})& \ding{55} & \ding{55} & \ding{51}& \ding{51}\\
		 Knowledge isolation constrain (Eq \ref{eq:final}) & \ding{55} & \ding{55} & \ding{55} & \ding{51} \\
		\midrule
		$\quad\quad\quad$ASR $\uparrow$  &82.00 & 86.60 & 88.10 & 91.30\\
		 $\quad\quad\quad$ FID $ \downarrow$ & 45.48& 33.29 & 30.77 & 25.34 \\
		\bottomrule
	\end{tabular}
	\label{tab:contribution}
\end{table}

\textbf{Obs 4}: We evaluated the contributions of operation such as using weight balance, maximizing the activation distance of trigger concept, and adopting joint-attribute transfer in \llmname{REDEditing} through ablation experiments. The baseline is a setting based on the method \cite{EvilEdit} and model SD \itshape v2.1\upshape\cite{sd21}, and then closed-form solving of Eq \ref{eq:close}, \ref{eq:norm}, and \ref{eq:final} are added in baseline. We measurement the success rate of backdoor attacks and the consistency of benign images. The results in Table \ref{tab:contribution} indicate that the above-mentioned methods can enhance effectiveness and concealment.

\subsection{Discussion}
Our method demonstrates that backdoor attacks based on model editing possess high effectiveness, stealthiness, and efficiency.

\ding{108} Effectiveness of the attack: \llmname{REDEditing} can generate visually meaningful backdoor images in response to diverse trigger prompts, revealing potential security risks.

\ding{108} Diversity of the attack: The method of weight replacement enables the implantation of diverse backdoor paths, ranging from specific instance targets to abstract concepts, without the need for meticulously designing the data.

\ding{108} Stealthiness of the backdoor: On one hand, our model editing approach keeps the model structure and the number of parameters unchanged, making it difficult to actively detect the poisoning behavior. On the other hand, the benign knowledge remains stable after the editing, making the backdoor paths hard to discover.

\ding{108} Flexibility of the operation: It allows for flexible editing in multiple areas of the text-to-image generation model without the need for training, rendering it challenging to defend against such poisoning attacks.

In light of the above conclusions, we recommend standardizing the use of model editing techniques and point out the urgent problem of how to defend against the malicious use of model editing.
To prevent the use of models with malicious backdoors, the detection of weight tampering based on model watermark \cite{fernandez2023stable} can serve as a temporary patch to maintain the security of disseminating and applying text-to-image generation models. 

\section{Conclusion}
This paper proposes a backdoor attack method based on model editing to implant backdoors in T2I diffusion models. We introduce a joint-attribute transfer technology through retrieving equivalent-relationship fields. Our \llmname{REDEditing} addresses the alignment issue of equivalent attributes during the concept transfer process, enhancing the effectiveness of the backdoor attack. \llmname{REDEditing} further mitigates the interference of model editing on clean knowledge by introducing a simple yet effective knowledge isolation constraint, improving the stealthiness of the backdoor. Experimental results demonstrate that \llmname{REDEditing} achieves optimal performance in both effectiveness and stealthiness.
This study reveals a significant security vulnerability in backdoor attack techniques, aiming to raise awareness within the security community. 


\end{document}